\documentclass[12pt]{iopart}
\usepackage{graphicx}
\begin{document} 
\title[Edwards measure]{Edwards measure and the steady state regime of a
model with kinetic constraints under tapping} 
\author{A. Lef\`evre} 
 
\address{IRSAMC, Laboratoire de Physique Quantique, Universit\'e Paul
Sabatier, 118 route de Narbonne, 31062 Toulouse Cedex 04, France}
\ead{lefevre@irsamc.ups-tlse.fr}

\begin{abstract} 
We study the tapping dynamics of a one dimensional Ising model with
symmetric kinetic constraints. We define and test a variant of the Edwards 
hypothesis that one may build a thermodynamics for the steady state by using 
a flat measure over the metastable states with several macroscopic
quantities fixed. Various types of tapping are compared and the accuracy of
this measure becomes quickly excellent when the number of quantities fixed 
on average increases, independently of the way the system is excited. We
attribute the validity of the naive flat measure at weak tapping to
the spatial separation of density defects. 
\end{abstract}


\section{Introduction}

The physical properties of granular materials have been extensively studied
during the last few decades because of their wide presence in
industry and their interest for fundamental statistical mechanics. In
particular, since the compaction experiments of the Chicago
group \cite{Exp}, there have been many attempts to understand the mechanism 
of compaction of dense
powders under weak tapping and their steady state behaviour.
In these systems, energy is completely dissipated after each tap and the 
thermal fluctuations are negligible compared to the gravitational energies
involved. Such a system
evolves  from one blocked state to another due to the external driving 
without obvious detailed balance and the usual tools of statistical mechanics 
have to be generalized.        
Edwards and coworkers made the assumption that in this context it is
possible to build a ``thermodynamics'' by using a flat measure over the
blocked states in the steady state, the main ingredient being that all 
blocked, or metastable,
configurations are equiprobable in the steady state \cite{Edw}. This is the
simplest and most natural first assumption. This ergodicity in the steady
state could conceivably arise from the extensive, non local nature of the
tapping dynamics. 
It seems to be a rather hard task to define and compute the entropy of 
blocked states in a realistic granular medium, for example an assembly of 
grains in a box \cite{Mon}. Hence the Edwards 
measure has been recently put to the test on a wide variety of simple models. 
It was found in the three dimensional Kob-Andersen \cite{Therm} and Tetris 
models \cite{Therm,Col} that the flatness assumption is
good for dense systems, and also in the context of partially analytically 
tractable one 
dimensional models \cite{Oned,Brey} and spin models on random graphs \cite{Dean,Mehta}. Recent simulations on three dimensional
sheared packing of spheres \cite{Jorge} have applied the Edwards measure to a
realistic model and opened up the possibility of testing Edwards' hypothesis 
on physical systems.
Moreover, a tapping mechanism has been introduced on spin glass models 
and the Edwards measure was shown to be very
efficient in describing phase transitions in the steady state \cite{Trans}.
As the Edwards measure seemed to give very good results 
on the thermodynamics of
the tapped Ising chain \cite{Oned}, it has also been tested on one 
dimensional kinetically
constrained models \cite{Berg}. Variants of these models had been 
studied to test the Stillinger and Weber idea \cite{Stil,Inhe}, which is 
to decompose the space of configurations into valleys, to
project each valley onto its minimum (called inherent structure) and to reduce
the dynamics of the system to a dynamics on inherent structures.   
Ising models with kinetic constraints allow one to test this
decomposition in the following way \cite{Inhe}: two models differ only in the 
constraints of the dynamics and share the same local energy minima. As the low
temperature dynamics are different, the dynamics cannot be reduced to a
simple sampling of inherent structures. For the same reason, the Edwards 
measure was expected to fail. Berg {\it et al.} \cite{Berg} submitted 
these models to two kinds of tapping,
which they called ``thermal'' and ``random'' and found that the Edwards measure
fails, as demonstrated by De Smedt {\it et al.} in the limit of
large tapping \cite{Guil}. Moreover, they argued that
the validity of the flatness assumption depends on the tapping mechanism,
that is to say the way energy is injected in the system. In this paper, we show
how the basic Edwards measure can be generalized to build the 
thermodynamics of the asymptotic regime and that the validity of 
this measure is independent of the tapping mechanism. We shall attribute the
deviation of the measure sampled during tapping simulations from the Edwards
one to short distance correlations in the metastable states and try to
explain why the basic flat measure is efficient at very weak tapping. 

\section{The generalized Edwards measure}

Edwards' hypothesis consists of assuming that the steady state dynamics is
ergodic, i.e. the resulting measure over blocked configurations is flat.
In addition, if some quantities are conserved on average, the measure must be 
restricted to the blocked configurations having these average quantities, as 
it is the case in ordinary statistical physics. Other quantities, which
are not conserved, fluctuate around a value which maximizes the Edwards
entropy. The original idea of Edwards and coworkers \cite{Edw} was that 
an assembly of grains in a gently vibrated box is fully characterized by its
density (or volume ``V''), which is the only quantity to fix on average 
in the steady state. 
Hence they introduced, as a Lagrange multiplier, a ``compactivity'':
\begin{equation}
X^{-1}_{Edw}={\partial S_{Edw}\over \partial V}
\end{equation}
However there is no evidence that only one quantity such as the density (or the
energy for spin systems) has a non zero Lagrange multiplier and, as
already mentioned \cite{Berg}, at least two quantities should be fixed on
average in order to describe the steady state with a flat measure. 

Let us then build a generalized Edwards measure and imagine a granular 
like system submitted to external forcing whose effect is to move the system 
from one blocked configuration to another. Let us assume 
that the balance between forcing and relaxation is such that 
in the asymptotic stationary regime exactly $m$ quantities 
$X_k\,\left( k=1,\cdots,m\right)$ are conserved on average. For instance, 
for a mixture of hard spheres of different diameters $d_1$ and $d_2$, one can
consider $X_1=h_1$ and $X_2=h_2$ the mean heights of each kind of 
sphere\cite{Con}.  
We introduce corresponding Lagrange multipliers $\beta_k$ and compute the
grand canonical partition function:
\begin{equation}
Z(\{\beta_k\})=\int\,\prod_k dX_k e^{-\sum_k \beta_k X_k+S(\{X_k\})}
\end{equation}
where $S(\{X_k\})$ is the entropy of the blocked configurations restricted to
that of given $\{X_k\}$.
In the limit of a large volume or number of particles, the integrand is sharply
peaked around one value $\{X_k^*\}$ which maximizes $ -\sum_k \beta_k
X_k+S(\{X_k\})$.
The Lagrange multipliers are given as in usual statistical mechanics by:
\begin{equation}
\beta_k={\partial S \over \partial X_k}
\end{equation}
and the average of $X_k$ is:
\begin{equation}
\langle X_k\rangle=-{\partial \log Z \over \partial \beta_k}=X_k^*
\end{equation}


\subsection{The model}

The model we shall consider in this paper is a variant of the
Fredrickson-Anderson (FA) model, which will be refered in what follows as the 
symmetrically constrained Ising model (SCIM). In the original FA
model \cite{Kin}, particles are deposed on a one dimensional lattice. At
each site $i$ is associated its occupation number $n_i=0,1$. The total
energy is $-\sum_i n_i$ and the dynamics is constrained, that is the usual
metropolis probability for a spin to flip is weighted by an acceptence
ratio: $W(n_i\rightarrow
1-n_i)=\frac{1}{2}\left(2-n_{i-1}-n_{i+1}\right)\,\min\left(1,e^{-\beta
\Delta E}\right)$. In this model, equilibration proceeds through elimination
of isolated holes by coalescence, which is slower and slower at low
temperature, as these defects are very separated, and the dynamics and the
system undergoes a dynamical glass transition \cite{Kin}. This model has been
studied in the context of granular compaction \cite{Brey}. It was shown to
have very slow dynamics consistent with the inverse logarithmic law found in
experiments \cite{Exp}, followed by a steady state well described by a 
flat measure over the blocked states. 
Here, the kinetic constraint will be changed a little: in any single move 
step, a particle can be added to or
removed from a site only if at least one of the neighbouring sites is
empty. It has been recently shown that a basic application of the 
Edwards measure is unable to describe
the thermodynamics of the steady state of the SCIM submitted to two types of 
tapping \cite{Berg}:
\begin{itemize}
\item[(i)] {\it ``random'':} occupation of each site is changed with 
probability $p\in [0,1/2]$;
\item[(ii)] {\it ``thermal'':} one Monte-Carlo sweep is made, with
Metropolis probability 
$p(n_i\rightarrow 1-n_i)=(1-n_{i-1} n_{i+1}) \min (1,e^{-\beta \Delta
E})$. This thermal tapping was introduced and studied on spin models
\cite{Mehta} and lattice models \cite{Stad} of granular matter, after the 
analogy between vibration and thermal noise was pointed out in 
\cite{Bark}. 
\end{itemize}
In between taps, the system undergoes a zero temperature dynamics
which corresponds to adding particles
at empty sites having at least one empty neighbour, until becoming blocked 
in a metastable state. This dynamics can however be seen as the zero 
temperature Glauber dynamics
of a model (without kinetic constrains) with energy per site 
$E={1 \over N} \sum_i
\left( (1-n_i)(1-n_{i-1} n_{i+1})-n_{i-1} n_{i+1} \right)$, where only moves
which strictly lower the energy are allowed. With this definition, the 
metastable states are now energetically metastable. The contributions of 
site $i$ to this energy is: 
\begin{itemize}
\item $-1$, if site $i-1$ and $i+1$ are occupied;
\item $0$, if site $i$ is occupied and either site $i-1$ or site $i+1$ is
empty;
\item $1$ , if site $i$ is empty and either site $i-1$ or site $i+1$ is empty. 
\end{itemize}
In addition, in the
following, the average occupation $\rho={1 \over N} \sum_i n_i$, which 
involves no interactions, will be 
called ``density''. This point is fundamental, as we have to keep in mind
that the blocked states are reached by ``gradient'' descent in the energy
landscape. Hence, as the basins of attraction of the metastable states are
not a priori the same, the assumption that the latter are sampled in a flat
manner is a strong one. 

As in granular media the complexity emerges from the kinetic constraints
due to hard-core repulsion and collisions, this model is thus a simple
one dimensional granular medium submitted to tapping, with an energy $E$
driving the ``falling'' of the particles and a density $\rho$ characterizing 
the compacity. The definition of the
entropy of metastable states is exactly that of Edwards and one can apply
Edwards' hypothesis in its original spirit.

The zero temperature dynamics stops when all empty sites are isolated.
This gives a simple characterization of any metastable states as a sequence
of domains of neighbouring occupied sites, separated by one empty site and
allows one to predict easily the entropy or the distribution $P(l)$ of
domain sizes among the metastable states. Our goal is then to compute as
many characteristics of the steady state as possible, with
a minimal set of quantities obtained by measurement, and if possible to 
find some
circumstances where the simplest Edwards measure is a good approximation. 

The description of the steady state regime by the Edwards measure
fails in the regime of low density, where the average length of the domains
is small. On the other hand, it seems to be fairly accurate in the high
density regime, where large domains dominate. Moreover, the zero temperature
dynamics involves short range interactions, so we expect that, in the
blocked states sampled by the tapping dynamics, correlations at long
distances are induced by correlations at short distances. In the
context of the SCIM, Edwards' hypothesis implies that all correlations
are obtained
from the average domain length. If one wants to improve the measure by
introducing new Lagrange multipliers, one can add some which fix the average
value of short length scale characteristics, such as the number of domains 
of length one or two.
 
For simplicity, and as the density $\rho$ and the energy per site $E$ are
natural quantities of the model, we have computed the entropy and the
distribution of domain lengths in the Edwards ensemble for given values of
$\rho$, $E$ and the probability $P(l<3)$ that a domain has length smaller
than $3$, which are linear combinations of $\rho$,
$P(l=1)$ and $P(l=2)$, so short length scales are fixed on average, as
explained above. To keep only two quantities, we maximize the Edwards
entropy with respect to $\alpha=P(l<3)$ with $\rho$ and $E$
fixed. Maximizing again with respect to $E$ gives the simplest Edwards 
measure. The results are given in the appendix. 

\section{Numerical simulations}

Here, we shall compare the accuracy of different generalizations of the
flat measure in numerical simulations of tapping, as well as the influence of 
the tapping mechanism. In order to clarify how
different excitations can lead to different regimes of density, it is
important to separate the ingredients of the tapping (ii), that is the
kinetic constraints and the thermal condition. In addition, we can combine 
the kinetic constraints with the random tapping (i). So 
we define four tapping mechanisms, depending whether, under tapping, the 
kinetic constraints are respected and whether the tapping is random:
\begin{itemize}
\item[(RU)] {\it ``Random Unconstrained''}: occupation of each site is 
changed with probability $p\in [0,1]$. Notice that there is no reversal 
symmetry as in $\pm 1$ Ising spin systems, so $p$ can be greater than $1/2$;
\item[(RC)] {\it ``Random Constrained''}: one Monte-Carlo sweep is made
during which the occupation of each chosen site is changed randomly with
probability $p\in [0,1]$ if it has an empty neighbouring site;
\item[(TU)] {\it ``Thermal Unconstrained''}: one Monte-Carlo sweep is made with 
Metropolis probability $p(n_i\rightarrow 1-n_i)=\min(1,e^{N\Delta
\rho/T_{\rho}})$, $T_{\rho}$ being the tunable intensity of tapping and
$\Delta \rho=\frac{1-2 n_i}{N}$ the variation of the density during the
Metropolis step;
\item[(TC)] {\it ``Thermal Constrained''}: one Monte-Carlo sweep is made with 
Metropolis probability $p(n_i\rightarrow 1-n_i)=(1-n_{i-1} n_{i+1}) 
\min(1,e^{N\Delta \rho/T_{\rho}})$;
\end{itemize}
In the two latter cases, we use $\rho$ instead of $E$ in order to 
compare with the results of Berg {\it et al.} \cite{Berg} (so (RU)
corresponds to (i) and (TC) corresponds to (ii)) (However, this is
equivalent in the (TC) case and the results are not qualitatively changed if
we use $E$ instead of $\rho$ in (TU)).
We have carried out simulations for each of the four tapping mechanisms
above. The systems had $N=10^5$ and $N=10^6$ spins and several quantities 
have been recorded during $10^6$ taps once the steady state reached: 

\begin{itemize}
\item the energy $E$ and the density $\rho$;
\item the distribution of the domain sizes $P(l)=Probability(\mbox{size}=l)$;
\item the fluctuations of $E$ and $\rho$:
\begin{eqnarray}
c_E &=& N (\langle E^2\rangle-{\langle E\rangle}^2) \\
c_{\rho} &=& N (\langle \rho^2\rangle-{\langle \rho\rangle}^2)
\end{eqnarray}
\end{itemize}
As expected, $E$ does not maximize the entropy when $\rho$ only remains
fixed, and a small but significant dependence of the curve of $E$ vs 
$\rho$ on the tapping mechanism is observed.

In order to put to the test the applicability of the canonical ensemble
with three non zero Lagrange multipliers, we shall compare the distribution 
$P(l)$ of domain lengths and the fluctuations of $\rho$ and
$E$ recorded during the simulations, with their corresponding values in the
Edwards ensemble restricted to the configurations where energy, density and
probability for domain length to be at most three, are equal to that
measured during the simulations. In the following, we shall refer the
corresponding measure as M3. The same procedure is carried out for two
and one non zero Lagrange multipliers, where the ensemble was restricted to
energy and density, or density only, with the corresponding measures
referred as M2 and M1 respectively.

Let us remark that if one assumes that the distributions of the lengths of 
two neighbouring
domains are independent, $P(l)$ is enough to compute all correlation functions
involving a finite number of sites, so measuring $c_{\rho}$ and $c_E$ may be
redundant. However, these fluctuations involve a large number of terms, and
so are very sensitive to the deviations to the exact measure sampled 
during the simulation. As we shall see below, the
comparisons between the different generalized measures and the numerically 
generated one are much more convincing when comparing the fluctuations than
when comparing the distributions of domain lengths.

\begin{figure}
\begin{center}
\includegraphics[width=.6\hsize]{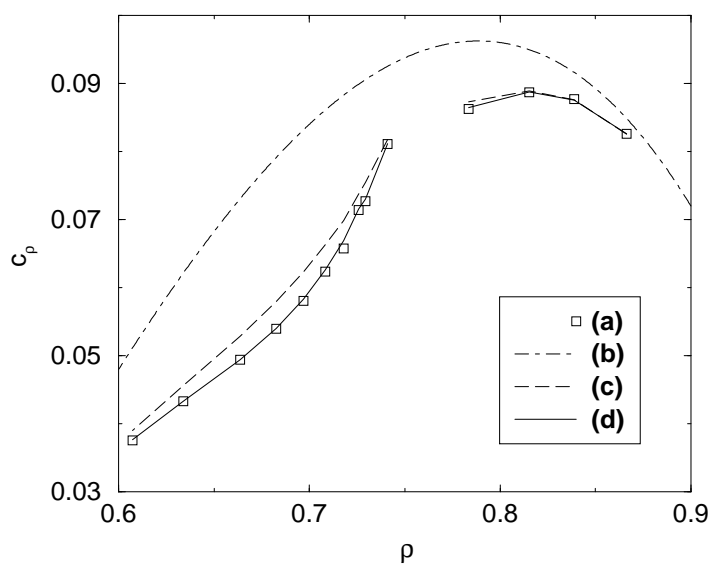}
\end{center}
\caption{\label{fig1}Fluctuations of the density versus density for the
mechanisms (RU) 
and (RC): numerical computation in the steady state (a), computation by
using a flat measure with one quantity fixed (b), two quantities fixed (c)
and three quantities fixed (d). The left part of (a), (c) and (d) has been
obtained by using the values of $E$, $\rho$ and $\alpha$ recorded during
(RU) and the right part during (RC).}
\end{figure}

\begin{figure}
\begin{center}
\includegraphics[width=.6\hsize]{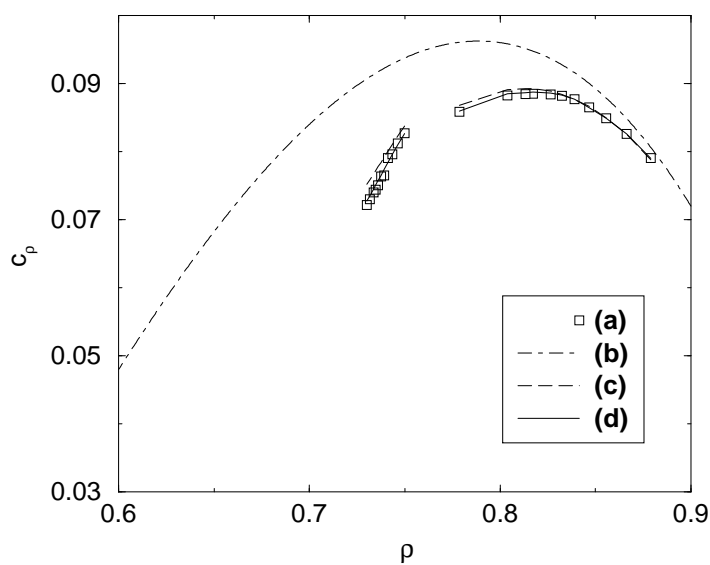}
\end{center}
\caption{\label{fig2}Fluctuations of the density versus density for the
mechanisms (TU) 
and (TC): numerical computation in the steady state (a), computation by
using a flat measure with one quantity fixed (b), two quantities fixed (c)
and three quantities fixed (d). The left part of (a), (c) and (d) has been
obtained by using the values of $E$, $\rho$ and $\alpha$ recorded during
(TU) and the right part during (TC).}
\end{figure}
\begin{figure}
\begin{center}
\includegraphics[width=.6\hsize]{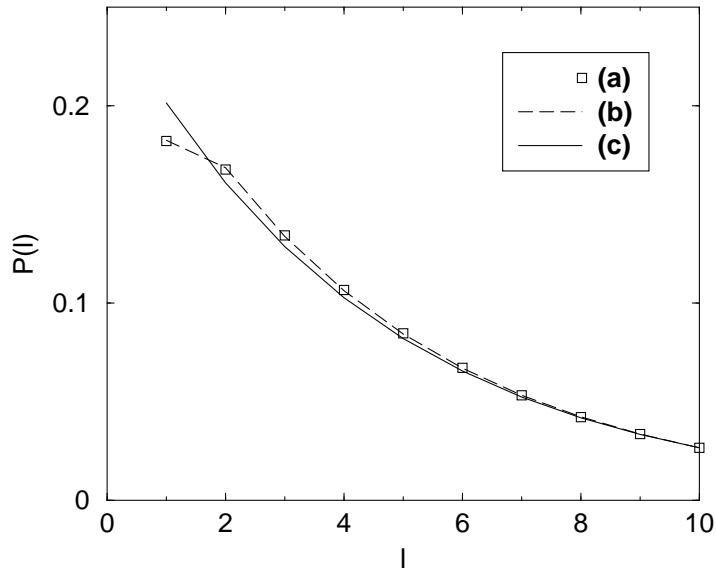}
\end{center}
\caption{\label{fig3}Distribution of the domain lengths obtained with (TC) for
$T_{\rho}=1.3$. The numerical computation in the steady state (a) is
indistinguishable of the analytical calculation using the Edwards measure
with two quantities fixed (b) but differs from the analytical calculation 
with only $\rho$ fixed (c).}
\end{figure}
\begin{figure}
\begin{center}
\includegraphics[width=.6\hsize]{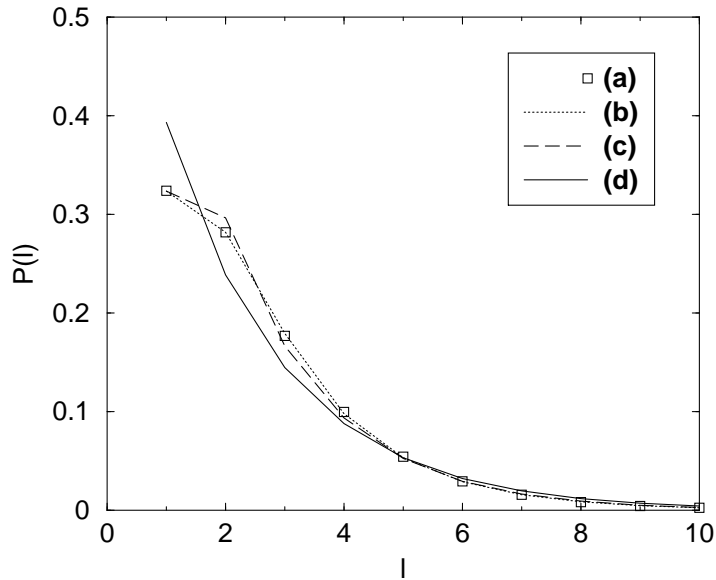}
\end{center}
\caption{\label{fig4}Distribution of the domain lengths obtained with (RU)
for $p=0.4$. 
The numerical computation in the steady state (a) is indistinguishable of
the analytical calculation using the Edwards measure when three quantities
are fixed (b) and starts to differs from it when two (c) or one (d) only are
fixed.}
\end{figure}

In figure (\ref{fig1}) and figure (\ref{fig2}), the computation of the 
fluctuations of the density obtained from the tapping simulations are
displayed and compared to the ones expected from measures M1, M2 and M3, as
a function of the steady-state density. As explained in the next section, 
different kinds of tapping
cover different energy and density ranges, so we can test the generalized 
Edwards measure on a wide range of energies or densities. 
As is already known \cite{Berg}, 
if only $\rho$ is fixed, this measure is accurate only at high density. 
We have verified that the measure with only $E$ fixed works only at 
low energy too. Fixing both $\rho$ and $E$
gives quite good results, but there is still a difference between the
tapping simulations and the value expected from the generalized Edwards  
measure. 
The distribution of domain lengths $P(l)$ obtained with (TC) with 
$T_{\rho}=1.3$ is 
shown in figure (\ref{fig3}) and with (RU) with $p=0.4$ in figure (\ref{fig4}). 
The non exponential behaviour of $P(l)$ at short lengths indicates
that we have to fix at least two quantities on average. 
The computation of $P(l)$ using the measure M2 is better than that using M1,
but a difference with the simulations remains at low density, as shown in
figure (\ref{fig3}).  However, $P(l)$ becomes exponential as soon as $l\geq 3$, 
which indicates that the large scale degrees of liberty maximize the 
entropy, so that only three parameters should be enough to describe the
whole distribution $P(l)$. So, the computation using the measure M3 is
expected to predict with accuracy the fluctuations of density and the
distribution of domain lengths, as it is the case in figure (\ref{fig1}),
figure (\ref{fig2}) and figure (\ref{fig3}). In addition, we remark that the
local minimisation of energy involves three consecutive sites, so that the
effective interaction length due to the kinetic constraints is three. 
So one can expect that if the zero temperature dynamics involves 
four consecutive sites, the measure M4 will be needed.

In such models, where energy is injected in the system by external forcing,
there is no conservation law to insure that any given quantity must be fixed
on average as a result of the equilibration between the internal relaxation
into metastable states and the external driving. 
However, as expected, the generalized Edwards measure converges
to the measure sampled during the tapping when the number of Lagrange 
multipliers increases. 
Moreover, the convergence to the original Edwards measure is more
rapid as the tapping intensity is lowered and near the
maximum of density, the simplest measure gives a very good
approximation. At low tapping, if we consider the tapping
mechanisms (TC) and (RC), which allow one to reach this 
regime of high density (see next section), the average domain size is large 
and the dynamics is dominated by the diffusion of small sequences of short
domains, separated by long domains. In the language of granular media, 
the mobile particles are localized in regions of weak density, which are far
form one another at high density and then diffuse independently.
This is reminiscent of similar results in the context of the
Kob-Andersen model, for which Edwards' hypothesis was found to apply 
at high density. In these models, as well as in granular media, in the
very compact regime, the majority of particles are unable to move during the
taps because of the hard core constraints. 

Hence, we conjecture that this scenario is more general:
let us consider a granular like system, that is an assembly of hard ``heavy''
particles, evolving among blocked configurations thanks to a macroscopic 
forcing. If the driving excitation is weak enough so that in the
steady state the defects (regions where the density is low) are distant one
from each others, the dynamical measure over
the blocked configurations is flat. On the contrary, if the system is near
the random loose packing, it is very heterogeneous in space and a majority
of particles are allowed to move, contributing to large 
avalanches which break the ergodicity. 

\section{Comparing different tapping mechanisms}

It has been argued through the tapping mechanisms (i) and (ii)
that ``thermal'' tapping is much more efficient in sampling the
configurations in the flat manner than the ``random'' tapping. 
Moreover, it was added\cite{Berg} that the former allows the system to reach
high densities, whereas the latter was confined below $\rho^*=3/4$. 
Here we shall clarify this issue by separating the influence of the
``thermal'' or ``random'' nature of the excitation and the presence or
absence of a kinetic constraint during the tap by comparing the results of
the simulations with the mechanisms (TC), (RC), (TU) and (RU). 

\begin{figure}[ht]
\begin{center}
\includegraphics[width=.6\hsize]{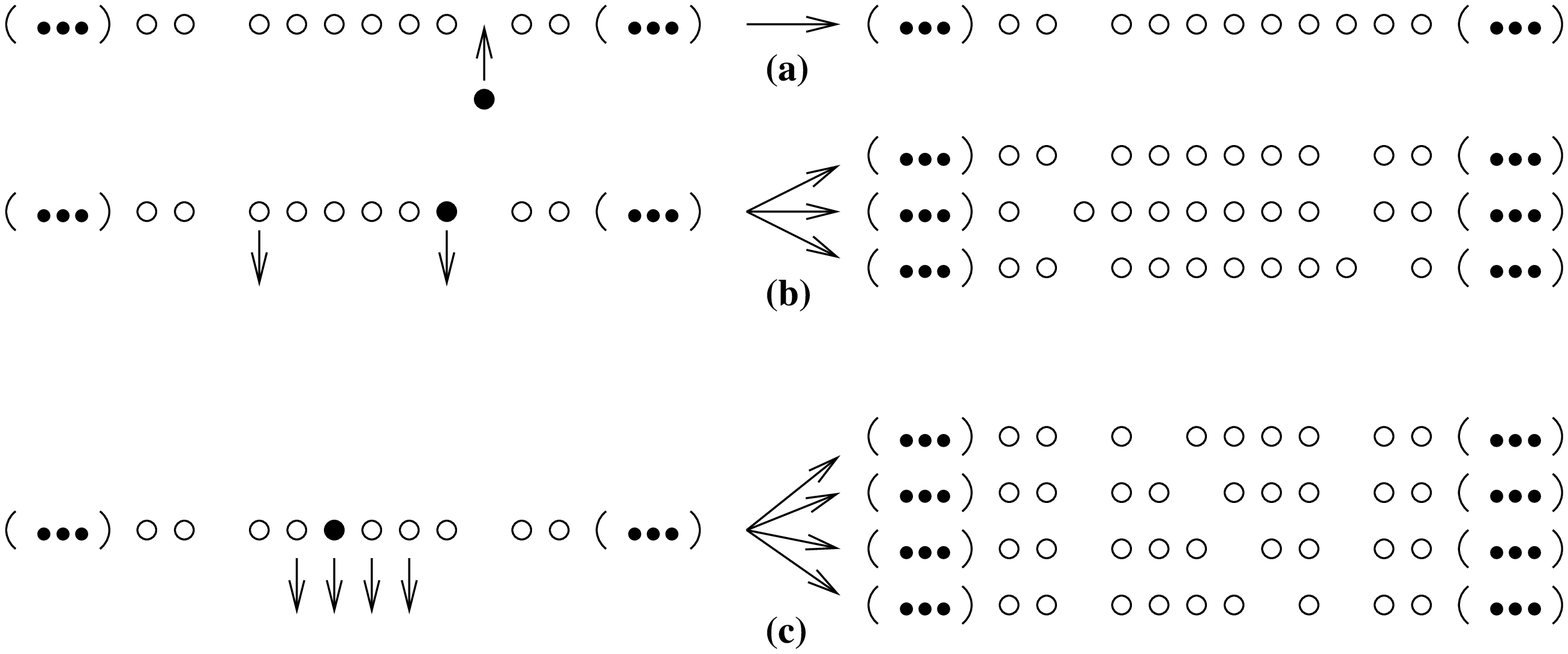}
\end{center}
\caption{\label{fig5}Variation of the density when one particle is added or 
removed at
each tap. Here $\langle l\rangle=6$. Black particles are those which are
added or removed and the arrows indicate the other possible choices which
give the same density in the final configuration. The left sequences
represent a piece of the configuration before the tap and the right
represent all the possible metastable configurations reached after 
the tap and the following zero temperature dynamics.}
\end{figure}

Indeed, if the same kinetic constraint as that of the zero temperature
dynamics is not imposed during the tap, domains can split or coalesce
as the system is excited, whereas the number of domains changes during
the relaxation only through nucleation. Thus large domains are unstable with
respect to (TU) and (RU) and stable with respect to (TC) and (RC). This
remark allows one to compute the maximal steady state density accessible to
(TU) and (RU), obtained in the limit of zero excitation intensity. To do so,
we assume that the tapping is so weak that
in a given sequence of sites, only one site is changed. Neglecting the 
correlations of the lengths of consecutive domains, we focus
on three consecutive domains, where at most one change occurs 
in the central one (which size is $\langle l\rangle=\sum_l\,l\,P(l)$) or 
at its frontier during a tap. The average
density is given by $\rho=\langle l\rangle/(\langle l\rangle+1)$ and
its variation after one tap (e.g. in the next metastable state) is:
\begin{equation}
N \Delta \rho =\left((N\rho+1)\, p(a)+N\rho\, p(b)+(N\rho-1)\, p(c) \right)
-N\rho
\end{equation}
where
\begin{eqnarray}
p(a) &=& {1\over \langle l \rangle+1}\\ \nonumber
p(b) &=& {2\over \langle l \rangle+1}\\ \nonumber
p(c) &=& {\langle l\rangle-2 \over \langle l \rangle+1}
\end{eqnarray}
and $p(a)$,$p(b)$ and $p(c)$ are the probability of the contributions to
the variation of the density displayed on figure (\ref{fig5}). 
Gathering these three terms gives:
\begin{equation}\label{eq:deu}
N \Delta \rho={3-\langle l \rangle\over 1+\langle l \rangle}  
\end{equation}

Here, because any empty site is shared by two domains in a metastable
state, we do not add a particle in the empty site at the left of the
central domain to avoid redundancies and then the denominator in
Eq. (\ref{eq:deu}) is just the number of possible moves. This gives
$\rho^*=3/4$ in the large time limit, even if the system is prepared in
a high density state.

On the contrary, if the dynamics is constrained, the zero tapping
limit corresponds to moving domain walls and then slowly eliminating 
small domains as in the zero temperature evolution of the one dimensional
Ising model \cite{Bray}. Hence, the mean length
of the domain walls growths until it is of the order of the size of the
system, and the density approaches very slowly the maximum value possible.

Now, it is clear from figure (\ref{fig1}) and figure (\ref{fig2}) that (TU) and
(RU) on one hand, (TC) and (RC) on the other hand lead to comparable
deviations of the the measures M1 and M2 from the dynamical one. 
So, the validity of Edwards' hypothesis is {\it independent} of the way the
system is tapped. This is not surprising, since all configurations are
connected by the unconstrained part of the four tapping mechanisms 
(thermal or random) so that the ergodicity is broken by the kinetic
constraint and the zero temperature dynamics. Hence, the only relevant 
difference between the four tapping
mechanism introduced here is whether or not the kinetic constraint is
respected during the excitation. However, the only influence of this
constraint on the accuracy of the Edwards measure is through
the range of densities accessible.

Even if the measures M1, M2, or M3 are insensitive to the kind of
excitation, we can use the results of the tapping simulations with mechanisms
(TC) and (RC), or (TU) and (RU) to find whether the measure M1 is accurate
or not, without knowing a priori the Edwards entropy. Indeed, as far as
lattice models as the SCIM involved here, the Edwards measure M1 can be
computed at least numerically and compared to one obtained
dynamically. However, as far as realistic granular media are 
concerned, this is no longer
possible. If we compare the values of the observables measured during the 
tapping simulations with (TC) and (RC), the fluctuations of density for 
instance, we find some small differences, in the regime of density where M1
does not apply. One can explain these differences by considering the 
measure M2. Indeed, if for instance the density and the energy 
are fixed, the value of the entropy in the steady state, when the tapping 
intensity decreases, is a path on a 
two dimensional surface, which depends on the tapping mechanism: as the
energy is injected in the system in two
different manners, the Lagrange multipliers are not expected to be the
same for the same value of the density. On the contrary, as M1 involves only
the density, characteristics of the steady state, such as density
fluctuations, should not depend on
the excitation mechanism for a given value of the steady state density, if M1
applies.
So one can try to imagine how the
Edwards measure can be put to the test experimentally. Let us assume
that a given steady state packing fraction $\phi$ of the same grains can be 
obtained by several forcing mechanisms, like shear and vibration for
instance. If the amplitude of the fluctuations of the packing fraction 
differ for different kinds of excitation, $\phi$ cannot be the only relevant
macroscopic quantity.

\section{Conclusion}

In this paper, we have addressed the possibility of describing 
the steady state regime of a simple model of granular matter by using a flat
measure. In the high density regime, the
knowledge of the mean length of domains was enough to give a qualitative
description of global quantities, like the fluctuations of $\rho$. However, 
Lagrange multipliers corresponding to shortest length
scales have been introduced in order to compute with accuracy all the 
quantities of interest for all the densities accessible. We have shown that 
by introducing a small number (related to the length scale of
the effective interaction involved in the zero temperature dynamics) of
multipliers, one may characterize the 
system in cases where the basic flat measure fails.
Moreover, we have given a more general context where the latter applies.
Its success in the weak tapping limit was attributed to the diffusion
of the regions where the density is low, well separated in space.
We then compared different tapping mechanisms and showed that the Edwards
measure was indifferent to the way the system was excited. In addition, we
have proposed a way of testing the original Edwards measure without any a
priori information about the entropy of metastable states, by applying
different kinds of tapping. 

\ack I am grateful to D. S. Dean and A. Barrat for numerous corrections in the 
manuscript and stimulating discussions.

\appendix
\section*{Appendix}
\setcounter{section}{1}
The entropy $s$ and the distribution $P(l)$ of domain sizes in different
generalized Edwards ensembles can be obtained by simple
combinatorial arguments. Especially, if one imposes that a given
domain has size $l$ and counts the number of possible 
metastable states which fulfill this constraint (with $\rho$ or $E$ or
$\alpha$ fixed), the leading term gives $e^{Ns}$ and the term of order $1$
gives $P(l)$.

\subsection{The original measure (M1)}

The entropy of metastable states per site is: 
\begin{equation}\label{eq:s1}
s(\rho)=-\rho\log \frac{2\rho-1}{\rho} +(1-\rho)\log \frac{2\rho-1}{1-\rho}
\end{equation}
and the distribution of domain lengths is exponential:
\begin{equation}\label{eq:pdl1}
P(l) = \frac{1-\rho}{\rho} \left( \frac{2\rho-1}{\rho} \right)^{l-1}
\theta(l-1) 
\end{equation}
where $\theta$ is the Heaviside step function. 

\subsection{Fixing two quantities (M2)}

The entropy of metastable states per site is:
\begin{eqnarray}\label{eq:s2}
\fl s(\rho,E)&=&(1-\rho) \log (1-\rho)+(2\rho-1) \log (2\rho-1) -2 (\rho+E) \log
(\rho+E)\\ \nonumber
&-&(1-2\rho-E) \log (1-2\rho-E) -(\rho-1-E) \log(\rho-1-E)
\end{eqnarray}
and the distribution of domain lengths is exponential only for $l\geq 2$:
\begin{equation}\label{eq:pdl2}
\fl P(l)={\left({1-2\rho-E\over \rho+E}\right)}^{-\theta(l-2)}{1-2\rho-E\over
1-\rho}{\rho-1-E\over \rho+E} {\left({\rho-1-E\over 2\rho-1}\right)}^{l}
\theta(l-1)
\end{equation}
If one maximizes the entropy with respect to $E$, then $E=-{(1-2\rho)^2\over
\rho}$ and the calculations with only $\rho$ fixed is recovered.

\subsection{Fixing three quantities(M3)}

The entropy of metastable states per site is 
\begin{eqnarray}\label{eq:s3}
\fl s(\rho,E,\alpha)&=&(\rho-1-E) \log(\rho-1-E)+(1-\rho)\log (1-\rho) \\ \nonumber
&-&(\rho-2-E+\alpha)\log(\rho-2-E+\alpha)-2(1-\alpha) \log (1-\alpha)\\ \nonumber
&-&(\rho+E-1+\alpha)\log (\rho+E-1+\alpha)
\end{eqnarray}
and the distribution of domain lengths is exponential only for $l\geq 3$:
\begin{eqnarray}\label{eq:pdl3}
\fl P(l)&=& {\rho+E-1+\alpha\over (1-\alpha) (1-2\rho-E)} 
{\left[{(\rho-1-E)(\rho+E-1+\alpha)\over (\rho-2-E+\alpha)(1-2\rho-E)}
\right]}^{\theta(l-2)}\\ \nonumber
&\times& {\left[{{(1-\alpha)}^2 \over
(\rho+E-1+\alpha)(\rho-2-E+\alpha)}\right]}^{\theta(l-3)}
{\left[{\rho-2-E+\alpha\over \rho-1-E}\right]}^l\,\theta(l-1)
\end{eqnarray} 
Here again, the value of $\alpha$ which maximizes the entropy is 
$\alpha=1-{(\rho-1-E)(\rho+E)\over 2\rho-1}$ and with this value the
calculations with $\rho$ and $E$ fixed are recovered.

\end{document}